\documentclass[aps,prl,twocolumn,superscriptaddress,prbib]{revtex4}

\usepackage{graphicx}%
\usepackage{dcolumn}
\usepackage{amsmath}
\usepackage{color}
\usepackage{multirow}
\usepackage{amssymb}

\makeatletter
\def\btt#1{\texttt{\@backslashchar#1}}%
\DeclareRobustCommand\bblash{\btt{\@backslashchar}}%
\makeatother

\newcommand{\bm}{\boldsymbol}

\begin{document}

\preprint{PREPRINT (\today)}

\title{Unsplit superconducting and time reversal symmetry breaking transitions in Sr$_2$RuO$_4$ under hydrostatic pressure and disorder. 	}

\author{Vadim Grinenko}
 \email{v.grinenko@ifw-dresden.de}
 \affiliation{Institute for Solid State and Materials Physics, Technische Universit\"{a}t Dresden, D-01069 Dresden, Germany}
 \affiliation{Leibniz-Institut f\"{u}r Festk\"{o}rper- und Werkstoffforschung (IFW) Dresden, D-01171 Dresden, Germany}

\author{Debarchan Das}
  \affiliation{Laboratory for Muon Spin Spectroscopy, Paul Scherrer Institut, CH-5232 Villigen PSI, Switzerland}

\author{Ritu Gupta}
  \affiliation{Laboratory for Muon Spin Spectroscopy, Paul Scherrer Institut, CH-5232 Villigen PSI, Switzerland}

\author{Bastian Zinkl}
  \affiliation{Institute for Theoretical Physics, ETH Zurich, CH-8093 Zurich, Switzerland}

\author{Naoki Kikugawa}
  \affiliation{National Institute for Materials Science, Tsukuba 305-0003, Japan}

\author{Yoshiteru Maeno}
  \affiliation{Department of Physics, Kyoto University, Kyoto 606-8502, Japan}

\author{Clifford W. Hicks}
  \affiliation{Max Planck Institute for Chemical Physics of Solids, D-01187 Dresden, Germany}
  \affiliation{School of Physics and Astronomy, University of Birmingham, Birmingham B15 2TT, United Kingdom }

\author{Hans-Henning Klauss}
 \affiliation{Institute for Solid State and Materials Physics, Technische Universit\"{a}t Dresden, D-01069 Dresden, Germany}

\author{Manfred Sigrist}
 \email{mansigri@ethz.ch}
  \affiliation{Institute for Theoretical Physics, ETH Zurich, CH-8093 Zurich, Switzerland}

\author{Rustem Khasanov}
\email{rustem.khasanov@psi.ch}
  \affiliation{Laboratory for Muon Spin Spectroscopy, Paul Scherrer Institut, CH-5232 Villigen PSI, Switzerland}

\begin{abstract}
There is considerable evidence that the superconducting state of Sr$_2$RuO$_4$ breaks time reversal symmetry. In the experiments showing time reversal symmetry breaking its onset temperature, $T_\text{TRSB}$, is generally found to match the critical temperature, $T_\text{c}$, within resolution. In combination with evidence for even parity, this result has led to consideration of a $d_{xz} \pm id_{yz}$ order parameter. The degeneracy of the two components of this order parameter is protected by symmetry, yielding $T_\text{TRSB} = T_\text{c}$, but it has a hard-to-explain horizontal line node at $k_z=0$. Therefore, $s \pm id$ and $d \pm ig$ order parameters are also under consideration. These avoid the horizontal line node, but require tuning to obtain $T_\text{TRSB} \approx T_\text{c}$.  To obtain evidence distinguishing these two possible scenarios (of symmetry-protected versus accidental degeneracy), we employ zero-field muon spin rotation/relaxation to study pure Sr$_2$RuO$_4$ under hydrostatic pressure, and Sr$_{1.98}$La$_{0.02}$RuO$_4$ at zero pressure. Both hydrostatic pressure and La substitution alter $T_\text{c}$ without lifting the tetragonal lattice symmetry, so if the degeneracy is symmetry-protected $T_\text{TRSB}$ should track changes in $T_\text{c}$, while if it is accidental, these transition temperatures should generally separate. We observe $T_\text{TRSB}$ to track $T_\text{c}$, supporting the hypothesis of $d_{xz} \pm id_{yz}$ order.

\end{abstract}

\maketitle


\section{Introduction}
For unconventional superconductors identifying the symmetry of the order parameter is crucial to pinpoint the origin of the superconductivity. Unconventional pairing states are distinguished from conventional ones by a non-trivial intrinsic phase structure which causes additional spontaneous symmetry breaking at the superconducting phase transition. This can lead, for instance, to a reduction of the crystal symmetry or the loss of time reversal symmetry. Indeed, several superconductors are known, which show experimental responses consistent with time reversal symmetry breaking (TRSB) superconductivity \cite{Heffner_PRB_1989, Jin_PRL_1992, Luke_PRL_1993, Riseborough_book_2002, Maisuradze_PRB_2010, Hillier_PRL_2012, Biswas_PRB_2013, Schemm_Science_2014, Shang_PRL_2018, Ghosh_JPCM_2021, Grinenko_Ba122_NatPhys_2020}.

TRSB superconducting states are formed by combining two or more order parameter components with complex coefficients. These components may be degenerate by symmetry, belonging to a single irreducible representation of the crystalline point group (as in the case of $p_x \pm ip_y$ or $d_{xz} \pm id_{yz}$ superconductivity on a tetragonal lattice), or they may come from different representations (for example, $d_{xy} \pm id_{x^2-y^2}$ superconductivity on a tetragonal lattice). In the following, we refer to the former as \textit{single-representation} and the latter as \textit{composite-representation} order parameters. For composite-representation order parameters, the two components will generally onset at different temperatures. The higher transition temperature becomes $T_\text{c}$, the superconducting critical temperature, and the lower temperature $T_\text{TRSB}$, the temperature where time reversal symmetry breaking onsets. The possibility of composite order parameters is usually dismissed out of hand, because it is unusual for two components that are not related by symmetry to be close enough in energy.
However, there are a few known examples: $s$ and $d_{x^2-y^2}$ are relatively close in energy in iron-based superconductors~\cite{Boehm_PRX_2014, Grinenko_Ba122_NatPhys_2020}, while both (U,Th)Be$_{13}$\cite{Riseborough_book_2002,Heffner_PRB_1989} and UPt$_3$\cite{Schemm_Science_2014,Luke_PRL_1993,Jin_PRL_1992} have split $T_\text{c}$ and $T_\text{TRSB}$.

Here, we study Sr$_2$RuO$_4$, an unconventional superconductor~\cite{Maeno_Nature_1994, Mackenzie_RMP_2003}, in which the origin of the superconductivity remains a mystery. Evidence that this superconductor breaks time reversal symmetry comes from zero-field muon spin rotation/relaxation (ZF-$\mu$SR) experiments \cite{Luke_Nature_1998} and polar Kerr effect measurements \cite{Xia_PRL_2006}. Phase-sensitive probes using a corner SQUID device give further support \cite{Nelson_Science_2004}. Moreover, the Josephson effect between a conventional superconductor and Sr$_2$RuO$_4$ reveal features compatible with the presence of superconducting domains, as expected for TRSB superconductivity \cite{Kidwingira_Science_2006, Nakamura_JPSJ_2012, Anwar_SciRep_2013}. For two decades the leading candidate state to explain these and other observations was the chiral $p$-wave state $p_x \pm ip_y$ (the lattice symmetry of Sr$_2$RuO$_4$ is tetragonal), which has odd parity and therefore equal spin pairing.  However, there is compelling evidence against an order parameter with such spin structure.  This evidence includes paramagnetic limiting for in-plane magnetic fields~\cite{Maeno_JPSJ_2012, Yonezawa_PRL_2013, Kittaka_PRB_2014} and the recently discovered drop in the NMR Knight shift below $T_\text{c}$~\cite{Pustogow_Nature_2019,Ishida_JPSJ_2020}. In combination with the above experimental support for TRSB superconductivity, this evidence compels consideration of $d_{xz} \pm id_{yz}$ order.

$d_{xz} \pm id_{yz}$ order would be a surprise because it has a line node at $k_z = 0$, which under conventional understanding requires interlayer pairing, while in Sr$_2$RuO$_4$ interlayer coupling is very weak. It has been proposed that $d_{xz} \pm id_{yz}$ order might be obtained through multi-orbital degrees of freedom; in this model the order parameter symmetry is encoded in orbital degrees of freedom, so interlayer pairing is not required~\cite{Suh_PRR_2020}. This form of pairing is also under consideration for URu$_2$Si$_2$~\cite{Kasahara_PRL_2007, Schemm_PRB_2015}. However, so far it has not been unambiguously confirmed in any material. To avoid horizontal line nodes, the composite-representation order parameters $s \pm id_{x^2-y^2}$ \cite{Romer_PRL_2019},  $s \pm id_{xy}$ \cite{Romer_PRB_2020} and $d_{x^2-y^2} \pm ig_{xy(x^2-y^2)}$ \cite{Kivelson_npjQuantMat_2020, Willa_PRB_2020} have also recently been proposed for Sr$_2$RuO$_4$. In contrast to $d_{xz} \pm id_{yz}$, these require tuning to obtain $T_\text{c} \approx T_\text{TRSB}$ on a tetragonal lattice.

In this work, to test whether the order parameter of Sr$_2$RuO$_4$ is of single- or composite-representation type we perform ZF-$\mu$SR measurements on hydrostatically pressurised Sr$_2$RuO$_4$ and on La-doped Sr$_{2-y}$La$_{y}$RuO$_4$. Both of these perturbations maintain the tetragonal symmetry of the lattice. If the order parameter has single-representation nature, $T_\text{TRSB}$ will therefore track $T_\text{c}$. If the order parameter is of the composite-representation kind, with $T_\text{TRSB}$ matching $T_\text{c}$ in clean, unstressed samples through an accidental fine tuning, then perturbations away from this point should in general split $T_\text{TRSB}$ and $T_\text{c}$, whether they preserve tetragonal lattice symmetry or not \cite{Zinkl_arxiv_2020}. Here, we have observed a clear suppression of $T_{\rm TRSB}$ at a rate matching the suppression
of $T_{\rm c}$. Our experimental results provide evidence in favour of single-representation nature of the order parameter in Sr$_2$RuO$_4$.

\section{Experimental design and Results}

\subsection{$\mu$SR on Sr$_2$RuO$_4$ under hydrostatic pressure}

\begin{figure}[tbh]
\includegraphics[width=0.95\linewidth]{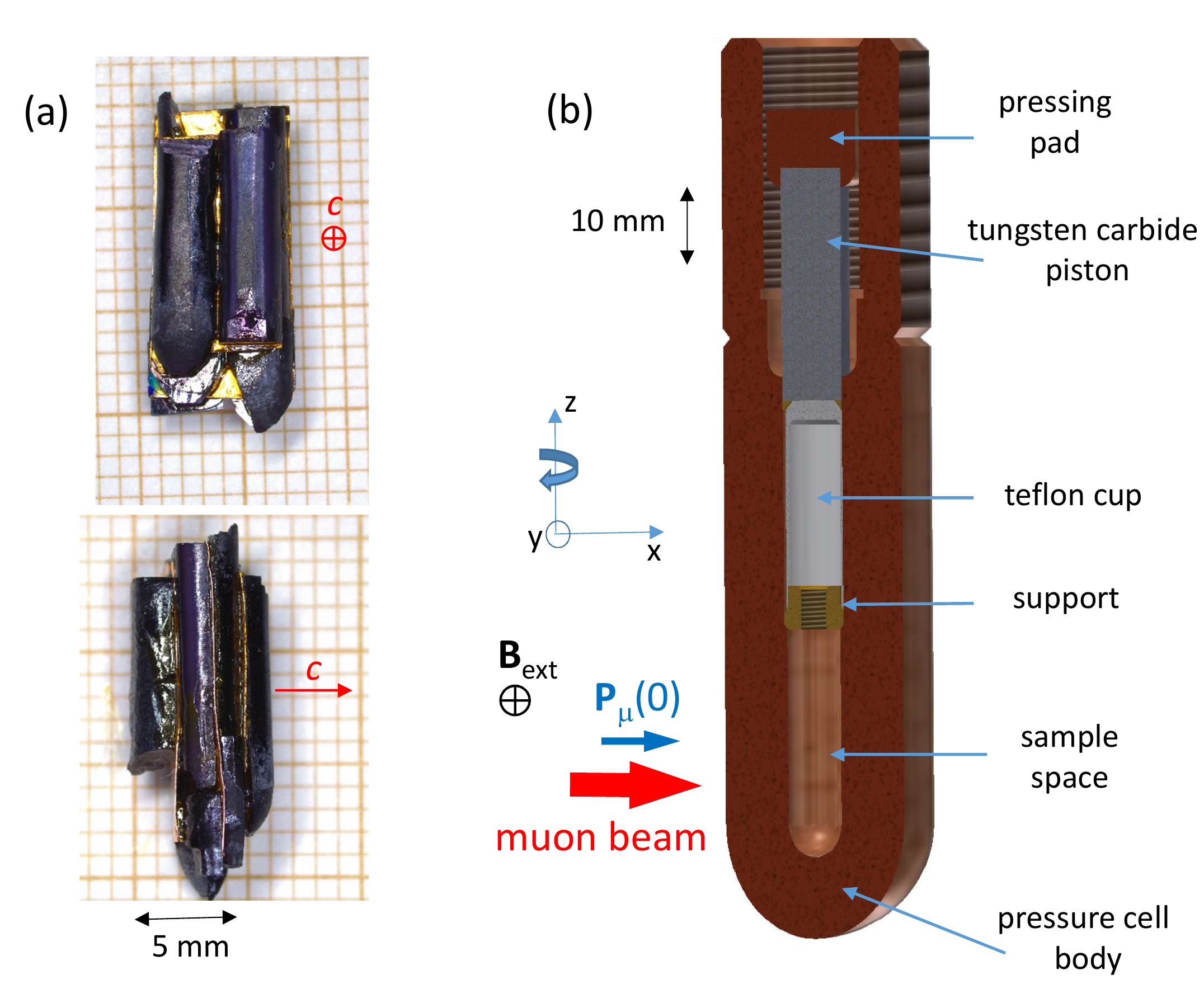}
\caption{Setup for hydrostatic pressure experiments. {\bf (a)} Sr$_2$RuO$_4$ sample, consisting of semi-cylindrical pieces glued on oxygen-free copper foils. The top and the bottom panels are the front and the side view, respectively. The crossed circle and the arrow indicate the orientation of the $c$-axis. {\bf (b)} Construction of the pressure cell \cite{Andreica_PhD-thesis_2001}. The sample and the pressure medium are surrounded only by beryllium-copper (the pressure cell body and the teflon cap support). The parts of the cell with strong $\mu$SR response (teflon cap and tungsten carbide  piston) are far from the sample and outside of the muon beam. The initial muon spin polarization ${\bf P}_{\rm \mu}(0)$ and the external field ${\bf B}_{\rm ext}$ in TF-$\mu$SR measurements are aligned along the $x$- and $y$-axes, respectively. By rotating the cell about the $z$-axis, the angle between $\mathbf{P}_\mu(0)$ and the sample $c$-axis can be varied.}
\label{fig:Sample-PressureCell}
\end{figure}

The hydrostatic pressure measurement setup is shown schematically in Fig.~\ref{fig:Sample-PressureCell}. Sr$_2$RuO$_4$ crystals of diameter $\varnothing\sim 3$~mm were affixed to oxygen-free copper foils, and assembled into an approximately cylindrical collection of total diameter $\varnothing \sim 7$~mm and total length $l\sim 12$~mm [see Fig.~\ref{fig:Sample-PressureCell}(a)]. The $c$-axes of the separate crystals were aligned to within $3^\circ$.

The pressure cell used in the present study [Refs.~\onlinecite{Andreica_PhD-thesis_2001, Khasanov_HPR_2016} and Fig.~\ref{fig:Sample-PressureCell}(b)] is a modification of a `classic $\mu$SR' clamped pressure cell~\cite{Khasanov_HPR_2016, Shermadini_HPR_2017}. It consists of a main body that encloses the sample and pressure medium, a teflon cap with a metallic support, a tungsten carbide piston, a pressing pad, and a clamping bolt (not shown) that holds the piston in place. All the metallic parts of the cell apart from the piston are made from a nonmagnetic beryllium-copper alloy, which is known to have a temperature-independent $\mu$SR response \cite{Khasanov_HPR_2016, Shermadini_HPR_2017, Andreica_PhD-thesis_2001}.  The main feature of this cell is that the only materials placed in the muon beam are the sample, the pressure medium, and this CuBe alloy. The muons had a typical momentum of 97~MeV/c, sufficient to penetrate the walls of the pressure cell.  The pressure medium was 7373 Daphne oil, which at room temperature solidifies at a pressure $p\approx$2.3~GPa~\cite{Murata_RSI_1997}. The maximum pressure reached here was 0.95~GPa, and therefore hydrostatic conditions are expected. The pressure was determined by monitoring the critical temperature of a small piece of indium (the pressure indicator) placed inside the cell with the Sr$_2$RuO$_4$ sample. Confirmation that essentially hydrostatic conditions were attained is provided by the fact that $T_\text{c}$ was observed to decrease linearly with pressure, whereas in-plane uniaxial stress on a GPa scale causes a strong non-linear increase in $T_\text{c}$~\cite{Steppke_Science_2017}.

\begin{figure*}[tbh]
\includegraphics[width=1.00\linewidth]{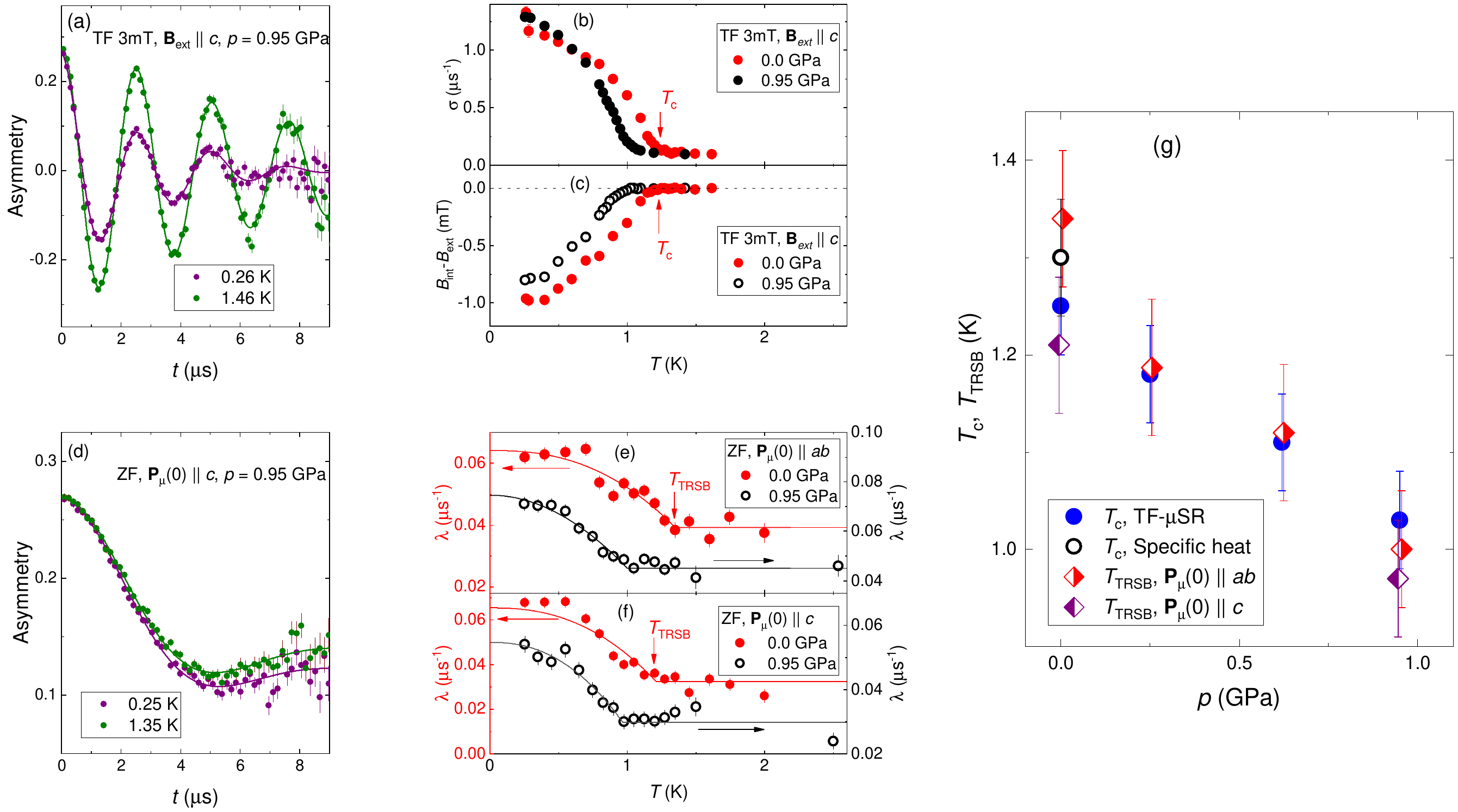}
\caption{Effect of pressure on $T_\text{c}$ and $T_\text{TRSB}$ in Sr$_2$RuO$_4$. {\bf (a)} TF-$\mu$SR time-spectra above and below $T_{\rm c}$ measured at $p=0.95$~GPa and $B_{\rm ext}= 3$~mT, with ${\bf B}_{\rm ext} \parallel c$. The plotted quantity is the detection asymmetry between two positron detectors, which is proportional to the muon spin polarisation $P_\mu(t)$. The solid lines are fits of Eq.~\ref{eq:P(t)}, with the sample and the pressure cell contributions described by Eqs.~\ref{eq:TF-P_sample} and \ref{eq:TF-P_pc}, respectively. {\bf (b)} and {\bf (c)} Temperature dependencies of the Gaussian relaxation rate $\sigma$ and the diamagnetic shift $B_{\rm int}-B_{\rm ext}\propto M_{\rm FC}$ at $p=0.0$ and 0.95~GPa. Arrows indicate the position of the superconducting transition temperature $T_{\rm c}$ at $p=0.0$~GPa.  {\bf (d)} ZF-$\mu$SR time-spectra above and below $T_\text{c}$, measured at $p=0.95$~GPa and with initial muon spin polarisation ${\bf P}_{\rm \mu}(0)\parallel c$. The solid lines are fits of Eq.~\ref{eq:P(t)}, with the sample and the pressure cell parts described by Eqs.~\ref{eq:ZF-P_sample} and \ref{eq:ZF-P_GKT}.  {\bf (e)} and {\bf (f)} Temperature dependencies of the ZF exponential muon spin relaxation rate $\lambda$ at $p=0.0$ and 0.95~GPa. In panel (e), $\mathbf{P}_\mu(0) \parallel ab$, and in panel (f), $\mathbf{P}_\mu(0) \parallel c$.  The solid lines are fits of Eq.~\ref{eq:lambda} to the data. Arrows indicate the position of $T_{\rm TRSB}$ at $p=0.0$~GPa. {\bf (g)} Dependence of $T_{\rm c}$ and $T_{\rm TRSB}$ on pressure. The displayed error bars for $\mu$SR data correspond to one standard deviation from the $\chi^2$ fit \cite{Hatlo_IEEE_2005}. The displayed error bars for $T_{\rm c}$ indicate the rounding of the transition on a scale of approximately 0.1~K. The error bars for $\mu$SR data and $T_{\rm TRSB}$ correspond to one standard deviation from the $\chi^2$ fit \cite{Hatlo_IEEE_2005}.  }
\label{fig:pressure_Tc-TRSB}
\end{figure*}

The samples used here were grown by the standard floating-zone method~\cite{SRO_sample-grow}. Measurements of heat capacity of pieces cut from the ends of the rods used here revealed an average $T_\text{c}$ of 1.30(6)~K (see Fig.~\ref{fig:SupplMat_specific-heat_SRO} and the Methods section), slightly below the limit of $T_\text{c}$ of 1.50~K for a pure sample.

$T_\text{c}$ and $T_\text{TRSB}$ were both obtained by means of $\mu$SR, ensuring that both quantities were measured for precisely the same sample volume. In the $\mu$SR method, spin-polarised muons are implanted, and their spins then precess in the local magnetic field. By collecting statistics of decay positrons in selected direction(s), the muon polarisation as a function of time after implantation, $P_{\rm \mu}(t)$, can be determined; the time-evolution of this polarisation is determined by the magnetic fields in the sample \cite{Yaouanc_book_2011}.

$T_\text{c}$ is determined through transverse-field (TF) measurements. An external field $B_\text{ext}$ of 3~mT was applied parallel to the crystalline $c$-axis and perpendicular to the initial
muon spin polarization $\mathbf{P}_\mu(0)$. Measurements were performed in the field-cooled (FC) mode. Details of the method and analysis are given in the Methods section.

Example TF-$\mu$SR time spectra at pressure $p = 0.95$~GPa, and at a temperature above $T_\text{c}$ and one below, are shown in Fig.~\ref{fig:pressure_Tc-TRSB}(a).  Above $T_\text{c}$, the spins of muons stopped in both the sample and the pressure cell walls precess with frequency $\omega = \gamma_\mu B_{\rm ext}$ (where $\gamma_\mu=2\pi \times 135.5$~MHz/T is the muon gyromagnetic ratio). The muon spin polarisation is seen to relax substantially on a 10~$\mu$s time scale. This is because approximately 50\% of muons are implanted into the CuBe, where the nuclear magnetic moments of Cu rapidly relax their polarisation.  Below $T_\text{c}$, the internal field in the sample becomes highly inhomogeneous due to the appearance of a flux-line lattice, and so the polarisation of the muons  that implanted in the sample also relaxes quickly.

TF-$\mu$SR measurements were performed at 0, 0.25, 0.62, and 0.95~GPa. Data at 0 and 0.95~GPa are shown in Fig.~\ref{fig:pressure_Tc-TRSB}, and at the other two pressures in Figs.~\ref{fig:p0p25_Tc-TRSB} and \ref{fig:p0p62_Tc-TRSB} in the Methods section.
Data are analysed as a sum of background and sample contributions, given by Eqs.~\ref{eq:TF-P_sample} and \ref{eq:TF-P_pc} (in the Methods section), respectively. From the sample contribution we extract a Gaussian relaxation rate, $\sigma$, and the diamagnetic shift of the field inside the sample, $B_\text{int} - B_\text{ext}\propto M_{\rm FC}$ \cite{Weber_PRB_1993} ($M_{\rm FC}$ is the field-cooled magnetization). Figures~\ref{fig:pressure_Tc-TRSB}~(b) and (c) respectively
show the temperature dependence of $\sigma$ and $B_\text{int} - B_\text{ext}$. $\sigma$ is given by $\sigma^2 = \sigma_\text{sc}^2 + \sigma_\text{nm}^2$, where $\sigma_\text{sc}$ and $\sigma_\text{nm}$ are the flux-line lattice and nuclear moment contributions, respectively. $\sigma_\text{sc} \propto \lambda_{ab}^{-2}$, where $\lambda_{ab}$ is the in-plane magnetic penetration depth; see Ref.~\onlinecite{Khasanov_PRB_2016} and the Methods section. The onset of superconductivity can be seen in both $\sigma$ and $B_\text{int}-B_\text{ext}$, as a transition rounded on a scale of approximately 0.1~K. The heat capacity measurements show a similar distribution of $T_\text{c}$'s; see Fig.~\ref{fig:SupplMat_specific-heat_SRO} and the Methods section.

The pressure dependence of $T_\text{c}$ is shown in Fig.~\ref{fig:pressure_Tc-TRSB}(g). The error bars in the figure are the rounding on the transition, and can be taken as an absolute error on $T_\text{c}$. When fitting $\sigma(T)$ and $B_{\rm int}(T)$ with model functions, the statistical error on the $T_\text{c}$'s extracted is considerably smaller, meaning that the error on changes in $T_\text{c}$ is low. A linear fit to $T_\text{c}(p)$ yields a slope $dT_\text{c}/dp = -0.24(2)$~K/GPa, which is in good agreement with literature  data~\cite{Shirakawa_PRB_1997, Forsythe_PRL_2002, Svitelskiy_PRB_2008}. The unpressurised $T_\text{c}$ is found to be 1.26(5)~K, in good agreement with 1.30(6)~K found in the heat capacity measurements, see the Methods section.

$T_\text{TRSB}$ is determined through zero-field (ZF) measurements. The signature of time reversal symmetry breaking is an enhancement in the muon spin relaxation rate below $T_\text{TRSB}$, indicating
the appearance of spontaneous magnetic fields. In these measurements, external fields were compensated to better than 2~$\mu$T, ruling out flux lines below $T_\text{c}$ as
the origin of this signal. An example of ZF-$\mu$SR time spectra above and below $T_\text{c}$, showing the faster relaxation below $T_\text{c}$, at $p = 0.95$~GPa is presented in
Fig.~\ref{fig:pressure_Tc-TRSB}(d). The pressure cell background is $T$-independent, so the increased signal decay comes from the sample.  The sample contribution was modelled by a two-component relaxation function: ${\rm GKT}(t)\cdot \exp(-\lambda t)$, in accordance with the results of Refs.~\onlinecite{Luke_Nature_1998, Luke_PhysB_2000, Maisuradze_PRB_2010, Hillier_PRL_2012, Shang_PRL_2018, Grinenko_SRO_NatPhys_2020}; see also the Methods section. Here, ${\rm GKT}(t)$ is the Gaussian-Kubo-Toyabe function describing the relaxation of muon
spin polarization in the random magnetic field distribution created by nuclear magnetic moments, and $\exp(-\lambda t)$ is a Lorentzian decay function accounting for appearance of spontaneous magnetic fields. Temperature dependencies of the exponential relaxation rate, $\lambda$, at 0 and 0.95~GPa, for independent measurements with the initial muon spin polarisation $\mathbf{P}_\mu(0) \parallel c$ and $\parallel ab$, are shown in Figs.~\ref{fig:pressure_Tc-TRSB}(e, f); ZF data at 0.25 and 0.62~GPa are shown in Figs.~\ref{fig:p0p25_Tc-TRSB} and
\ref{fig:p0p62_Tc-TRSB} in the Methods section.

To extract $T_\text{TRSB}$, $\lambda(T)$ is fitted with the following functional form:
\begin{equation}
   \lambda(T) =
\begin{cases}
\lambda_0  , & T>T_{\rm TRSB} \\
\lambda_0+\Delta\lambda\left[1-\left(\frac{T}{T_{\rm TRSB}}\right)^n\right] , & T<T_{\rm TRSB}
\end{cases}.
 \label{eq:lambda}
\end{equation}
$\lambda_0$ is the relaxation rate above $T_\text{TRSB}$, and $\Delta \lambda$ is the enhancement due to spontaneous magnetic fields. Where data were obtained both for $\mathbf{P}_\mu(0) \parallel c$ and
$\parallel ab$, the exponent $n$ is constrained to be the same for both polarisations. $T_\text{TRSB}$, $\lambda_0$, and $\Delta \lambda$ were obtained independently
for each pressure and muon spin polarisation. The resulting values of $T_\text{TRSB}$ are plotted in Fig.~\ref{fig:pressure_Tc-TRSB}(g).

Our ZF data yield the following three results: \\
(i) Where data were taken both for $\mathbf{P}_\mu(0) \parallel c$ and $\parallel ab$ (that is, at 0 and 0.95~GPa), $T_\text{TRSB}$ and $\Delta \lambda$ were found to be the same within resolution for both polarisations. [At 0~GPa, $\Delta \lambda = 0.027(4)$ and $0.033(3)$~$\mu{\rm s}^{-1}$, and at 0.95~GPa, $0.030(4)$ and $0.025(3)$~$\mu{\rm s}^{-1}$, for $\mathbf{P}_\mu(0) \parallel ab$ and $\mathbf{P}_\mu(0) \parallel c$, respectively.] This agrees with the zero-pressure results of Luke \textit{et al.} \cite{Luke_Nature_1998}. Because $\Delta \lambda$ reflects fields perpendicular to $\mathbf{P}_\mu(0)$, this result indicates that the spontaneous fields have no preferred orientation.\\
(ii) $\Delta \lambda$ was found to be pressure-independent within resolution  (including all pressures investigated: 0, 0.25, 0.62, and 0.95 GPa), having an average value of $\Delta \lambda = 0.026(2)$~$\mu$s$^{-1}$. This value corresponds to a characteristic field strength $B_\text{TRSB} = \Delta \lambda / \gamma_\mu = 0.031(2)$~mT. $B_\text{TRSB}$ has been found to vary from sample to
sample~\cite{Grinenko_SRO_NatPhys_2020}, and this value is in line with previous reports~\cite{Luke_Nature_1998, Luke_PhysB_2000, Shiroka_PRB_2012, Higemoto_JPS-Conf_2014}. \\
(iii) A linear fit yields $T_{\rm TRSB}(p)=1.27(3){\rm ~K}- p\cdot 0.29(5){\rm ~K/GPa}$. In other words, within resolution the rate of suppression of $T_\text{TRSB}$ under hydrostatic pressure matches that of $T_\text{c}$.\\

\subsection{$\mu$SR on Sr\textsubscript{1.98}La\textsubscript{0.02}RuO\textsubscript{4}}

\begin{figure*}[tbh]
	\centering
	\includegraphics[width=1.00\linewidth]{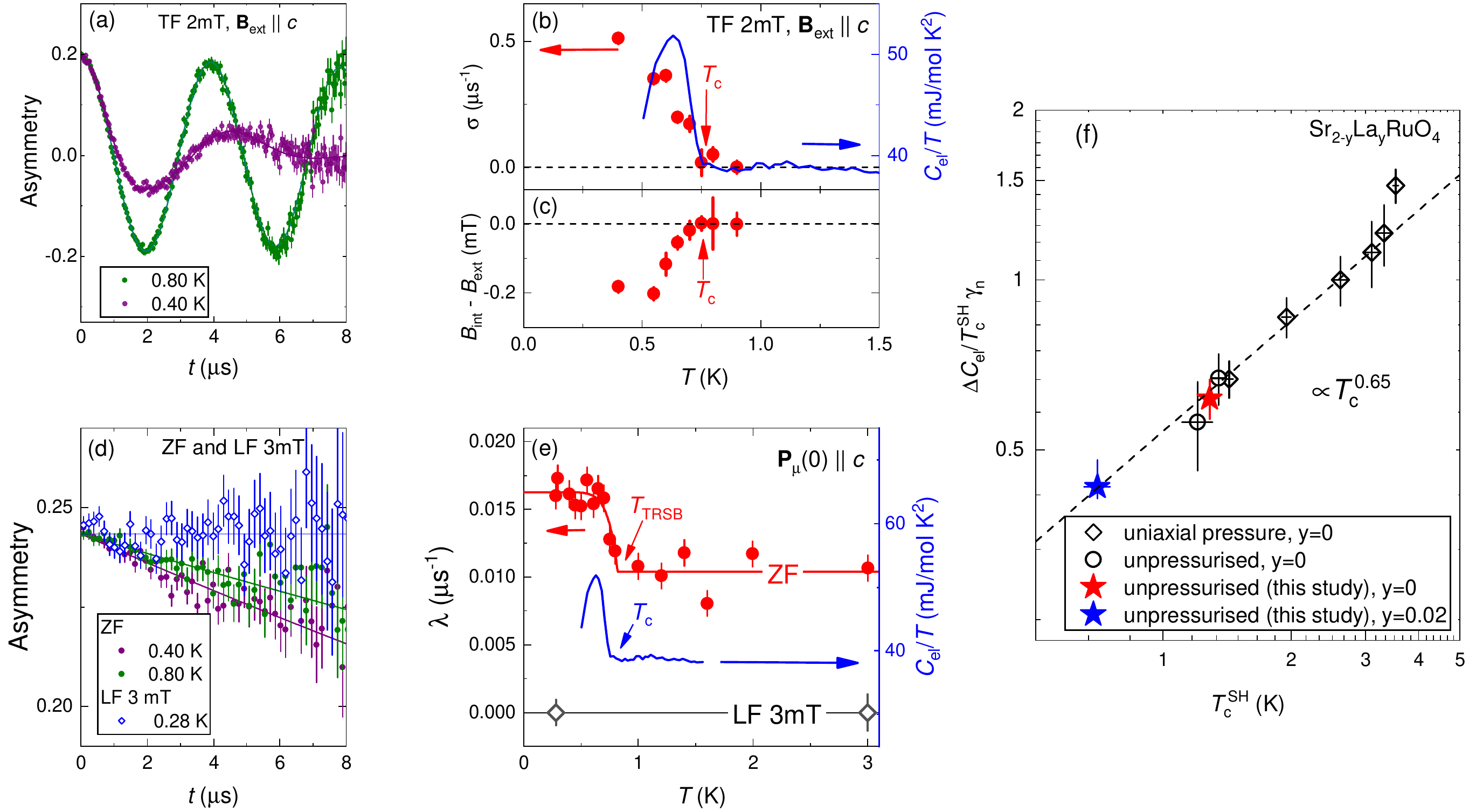}
\caption{TRSB in Sr$_{\rm 1.98}$La$_{\rm 0.02}$RuO$_4$. {\bf (a)} TF-$\mu$SR time-spectra above and below $T_{\rm c}$ measured at $B_{\rm ext}= 2$~mT with ${\bf B}_{\rm ext}\parallel c$. The solid lines are fits of Eq.~\ref{eq:TF-P_sample} to the data. {\bf (b)} and {\bf (c)} Temperature dependencies of the Gaussian relaxation rate $\sigma$ and the diamagnetic shift $B_{\rm int}-B_{\rm ext}$, respectively. Arrows indicates the superconducting transition temperature $T_{\rm c}$, determined from the TF-$\mu$SR data. The blue curve in panel (b) is the electronic specific  heat $C_{\rm el}/T$, measured on a small piece cut from the $\mu$SR sample. {\bf (d)} ZF- and LF-$\mu$SR time-spectra. ZF data from above and below $T_\text{c}$, measured with $\mathbf{P}_\mu(0) \parallel c$, are shown. The LF data are from $T$ well below $T_\text{c}$, and with $\mathbf{B}_\text{ext} = 3$~mT $\parallel \mathbf{P}_\mu(0)$. The solid lines are fits of Eq.~\ref{eq:LF-P_sample}. {\bf (e)} Temperature dependence of the ZF and LF exponential relaxation rate $\lambda$. The solid red line is the fit of Eq.~\ref{eq:lambda} to ZF $\lambda(T)$ data. The blue curve is, again, $C_{\rm el}/T$. Arrows indicates positions of $T_{\rm c}$ and $T_{\rm TRSB}$.  {\bf (f)} Double logarithmic plot of the normalized specific heat jump $\Delta C_{\rm el}/\gamma_{\rm n}T_{\rm c}^{\rm SH}$ versus $T_{\rm c}^{\rm SH}$ [$\gamma_{\rm n}$ is the Sommerfeld coefficient and $T_{\rm c}^{\rm SH}$ is the
transition temperature determined from $C_{\rm el}/T(T)$ by means of equal-entropy construction, see Fig.~\ref{fig:SupplMat_specific-heat_SRO}(a)].
Filled symbols: data from this work; open symbols: data taken from Refs.~\onlinecite{Grinenko_SRO_NatPhys_2020, Li_arxiv_2019}. The displayed error bars for $\mu$SR data correspond to one standard deviation from the $\chi^2$ fit \cite{Hatlo_IEEE_2005}. The error bars for $\Delta C_{\rm el}/\gamma_{\rm n}T_{\rm c}^{\rm SH}$ and $T_{\rm c}^{\rm SH}$  indicate uncertainty in selecting the temperature range for linear fit below $T_{\rm c}$. }
	\label{fig:La-doped}
\end{figure*}

Substitution of La for Sr adds electrons to the Fermi surfaces; in Sr$_{2-y}$La$_y$RuO$_4$ this doping drives the largest Fermi surface through a Lifshitz transition from an electron-like to a hole-like geometry, at $y \approx 0.20$~\cite{Kikugawa_PRB_2004, Shen_PRL_2007}. At $y=0.02$, the change in Fermi surface structure is minimal, and the main effect of the La substitution is to suppress $T_\text{c}$, through the added disorder. Heat capacity data, measured on a small piece cut from the $\mu$SR sample, give $T_\text{c} = 0.70(5)$~K, where the error reflects the width of the transition (see Fig.~\ref{fig:SupplMat_specific-heat_LaSRO}).

This sample was studied at zero pressure. With no pressure cell material in the beam, the background is much smaller. The typical muon momentum was 28~MeV/c, giving of approximately 0.1~mm implantation depth \cite{Yaouanc_book_2011}. Representative TF-$\mu$SR time spectra above and below $T_\text{c}$, where the applied field is $B_\text{ext} = 2$~mT parallel to the crystalline $c$-axis, are shown in Fig.~\ref{fig:La-doped}(a).  Below $T_\text{c}$, the muon spin polarisation relaxes almost completely on a 10~$\mu$s timescale, showing that essentially the entire sample volume is superconducting. The TF Gaussian relaxation rate $\sigma$ is shown in panel (b), and $B_\text{int}-B_\text{ext}$ in panel (c). These measurements yield $T_\text{c} = 0.75(5)$~K. The heat capacity data are also shown in panel (b).

ZF-$\mu$SR data are presented in Figs.~\ref{fig:La-doped}(d) and (e). Fitting with Eq.~\ref{eq:lambda} returns $\Delta \lambda = 0.007(1)$~$\mu$s$^{-1}$ and $T_\text{TRSB} = 0.8(1)$~K. This $\Delta \lambda$ is noticeably smaller than that obtained from the undoped Sr$_2$RuO$_4$ sample, corresponding to an internal field $B_\text{TRSB} \approx 0.01$~mT. It is, however, within the range of previous results~\cite{Grinenko_SRO_NatPhys_2020}. In qualitative agreement with data on a lower-$T_\text{c}$ Sr$_2$RuO$_4$, reported in Ref.~\onlinecite{Luke_PhysB_2000}, though here with more data at $T>T_\text{c}$ to be certain of the base relaxation rate,  this low value of $\Delta \lambda$ shows that $B_\text{TRSB}$ is not straightforwardly related to defect density.  At present, the origin of the sample-to-sample variation in $B_\text{TRSB}$ is unknown.

Longitudinal-field (LF) measurements can be employed to determine whether internal fields are static or fluctuating. If $B_\text{TRSB}$ is static, under an applied field parallel to $\mathbf{P}_\mu(0)$ that is considerably larger than $B_\text{TRSB}$, muon spin precession is greatly restricted and the spin polarisation does not relax (\textit{i.e.} the muon spins decouple from $B_\text{TRSB}$). In contrast, fluctuating $B_\text{TRSB}$ can still relax the muon spin polarisation \cite{Yaouanc_book_2011}. Data shown in panels (d) and (e) indicate that $\mathbf{B}_\text{ext} || \mathbf{P}_\mu (0)= 3$~mT fully suppresses the muon spin relaxation, and therefore that $B_\text{TRSB}$ is static on a microsecond time scale, in agreement with data on clean Sr$_2$RuO$_4$ reported in Ref.~\onlinecite{Luke_Nature_1998}. We note that LF measurements were not performed on the hydrostatically pressurised sample because the decoupling field for the Cu background is of the order of 10~mT, considerably stronger than that for Sr$_2$RuO$_4$.

\begin{figure*}[tbh]
\includegraphics[width=0.55\linewidth]{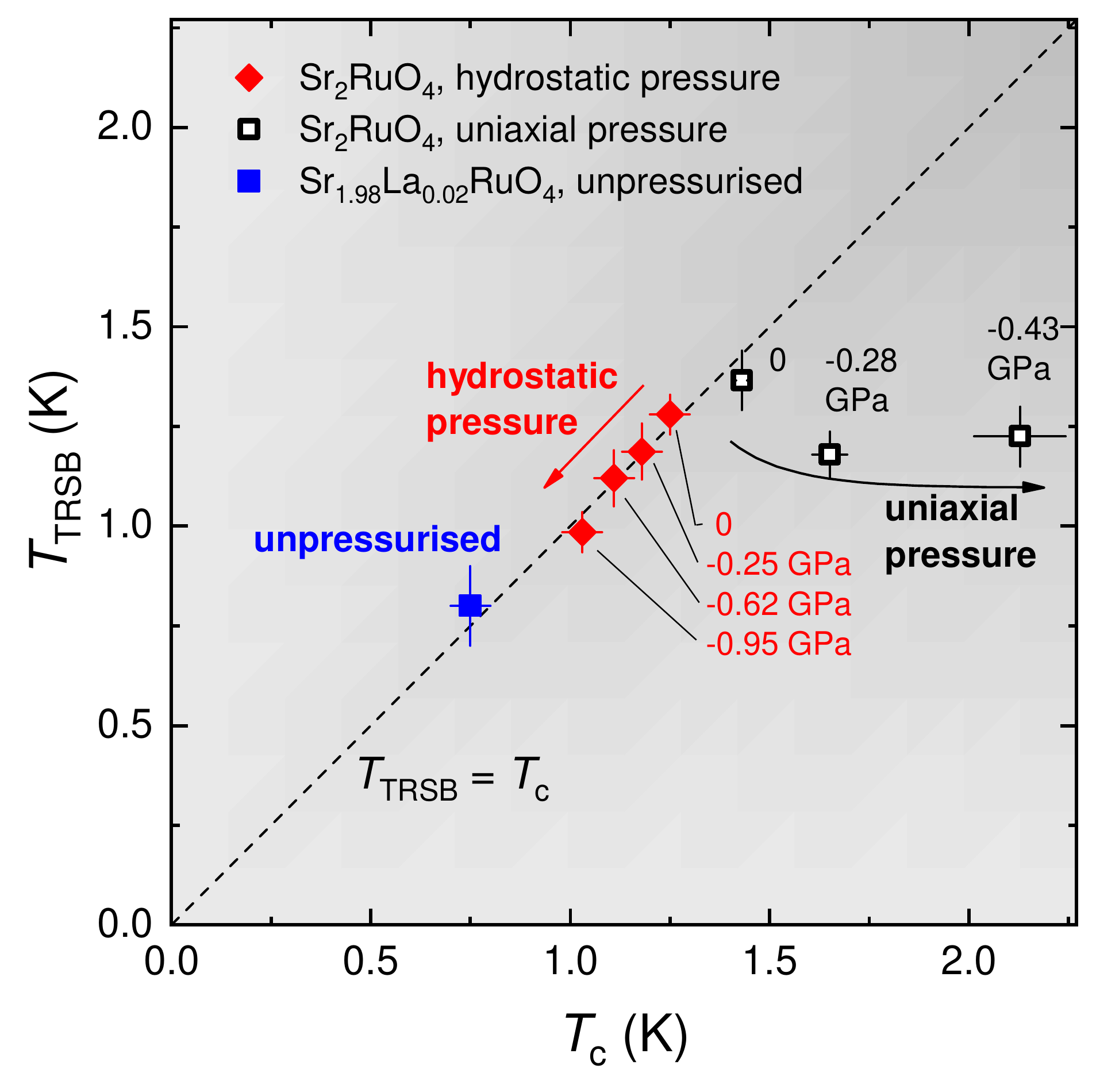}
\caption{Relation between $T_{\rm TRSB}$ and $T_{\rm c}$.  Dependence of the time reversal symmetry breaking temperature $T_{\rm TRSB}$ on the superconducting transition temperature $T_{\rm c}$. The
closed symbols correspond to the results obtained in present studies under hydrostatic pressure up to 0.95~GPa in pure Sr$_2$RuO$_4$ (diamonds) and in the La doped Sr$_{2-y}$La$_y$RuO$_4$ with $T_{\rm c}
= 0.75(5)$~K (square). The open squares are the uniaxial pressure data for undoped Sr$_2$RuO$_4$  from Ref.~\onlinecite{Grinenko_SRO_NatPhys_2020}. The dashed line corresponds to $T_{\rm TRSB}=T_{\rm c}$. The minus signs at the pressure values denote the effect of 'compression' of the sample volume. }

\label{fig:Tc-TTRSB_correlations}
\end{figure*}

\subsection{Heat capacity measurements}

The specific heat measurements  were performed at ambient pressure for several pieces of Sr$_{\rm 2-y}$La$_{\rm y}$RuO$_4$ single crystals. The results are presented in Figs.~\ref{fig:La-doped}~(b) and (e) for Sr$_{1.98}$La$_{0.02}$RuO$_4$ ($y=0.02$) and in the Methods section for Sr$_{2}$RuO$_4$ ($y=0$, Fig.~\ref{fig:SupplMat_specific-heat_SRO}), respectively. The specific heat jumps at $T_{\rm c}$ ($\Delta C_{\rm el}/\gamma_{\rm n}T_{\rm c}$, $\gamma_{\rm n}$ is the Sommerfeld coefficient) were further obtained in a way presented in Fig.~\ref{fig:SupplMat_specific-heat_LaSRO}.

Figure~\ref{fig:La-doped}~(f) summarises the $\Delta C_{\rm el}/\gamma_{\rm n}T_{\rm c}^{\rm SH}$ vs. $T_{\rm c}^{\rm SH}$ data for our Sr$_{\rm 2-y}$La$_{\rm y}$RuO$_4$  samples. Here $T_{\rm c}^{\rm SH}$ denotes the superconducting transition temperature determined from $C_{\rm el}/T$ vs. $T$ measurement curves by means of equal-entropy construction algorithm, see Fig.~\ref{fig:SupplMat_specific-heat_SRO}(a) and the Methods section. In addition, we have also included some literature data for Sr$_2$RuO$_4$ with different amount of disorder \cite{Grinenko_SRO_NatPhys_2020} and for Sr$_2$RuO$_4$ under uniaxial strain \cite{Li_arxiv_2019}. In total, Fig.~\ref{fig:La-doped}(f) compares Sr$_2$RuO$_4$ samples with a factor of five variation in $T_{\rm c}$. Remarkably, $\Delta C_{\rm el}/\gamma_{\rm n}T_{\rm c}$ vs. $T_{\rm c}$ data points scale as $T_{\rm c}^\alpha$ with $\alpha \approx 0.65$, which is distinctly different from the BCS behavior, where $\alpha = 0$ ($\Delta C_{\rm el}/\gamma_{\rm n}T_{\rm c} = const$).
Just a single point at $T_{\rm c}\simeq 3.5$~K deviates from the scaling behavior, which might be associated with tuning the electronic structure of Sr$_2$RuO$_4$ close to a van Hove singularity \cite{Li_arxiv_2019}.
The results presented in Fig.~\ref{fig:La-doped}~(f) indicate, therefore, that the perturbation changes the gap structure on the Fermi surface, i.e. its 'anisotropy' or the distribution among the three different bands which can lead to a renormalization of the specific heat jump being not simply proportional to normal state specific heat at $ T_{\rm c} $. Surprisingly, both hydrostatic pressure and La-doping collapse on the same curve here. The fact that this scaling is monotonic shows that both perturbations do not yield a strong qualitative change of the gap anisotropy.

It is worth noting here, that for most Fe-based superconductors, $\Delta C_{\rm el}/\gamma_{\rm n}T_{\rm c}$  follows approximately the  BNC (Bud'ko-Ni-Canfield) scaling behavior $\Delta C_{\rm el}/\gamma_{\rm n}T_{\rm c} \propto T_{\rm c}^\alpha$ with $\alpha \approx 2$ \cite{Budko_PRB_2009}, which is considered to be a consequence of the unconventional multiband $s_{\rm \pm}$ superconductivity. The change of the superconducting pairing state in the Ba$_{\rm 1-x}$K$_{\rm x}$Fe$_2$As$_2$ system results in abrupt change of the scaling behavior leading to an intermediate $s+is$ state \cite{Grinenko_Ba122_NatPhys_2020}.

\section{Discussion}

In a previous ZF-$\mu$SR experiment, in-plane uniaxial pressure, which does lift the tetragonal symmetry of the unpressurised lattice, was found to induce a strong splitting between $T_\text{c}$ and $T_\text{TRSB}$~\cite{Grinenko_SRO_NatPhys_2020}. Uniaxial pressure drives a strong increase in $T_\text{c}$, while $T_\text{TRSB}$ varies much more weakly, probably decreasing slightly with initial application of pressure.  The microscopic mechanism yielding the signal observed at $T_\text{TRSB}$, a weak enhancement in muon spin relaxation rate, remains unclear: the main proposed mechanism, magnetism induced at defects and domain walls by a TRSB superconducting order, is unproved experimentally~\cite{Kirtley_PRB_2007, Curran_PRB_2014}. At present, the link between enhanced muon spin relaxation and TRSB superconductivity is, therefore, mainly empirical, based on:
(i) the facts that it is a signal seen in only a small fraction of known superconductors, that it generally appears at $T_\text{c}$, and
(ii) the general notion that TRSB superconductivity can in principle generate magnetic fields while muons detect magnetic fields.
In Ref.~\onlinecite{Grinenko_SRO_NatPhys_2020}, careful checks were performed to rule out instrumentation artefact as the origin of the signal at $T_\text{TRSB}$, and it was further argued that this signal is extremely difficult to obtain from a purely magnetic mechanism. Nevertheless, the weak observed variation of $T_\text{TRSB}$, while $T_\text{c}$ varied strongly, raised some doubt as to whether this signal is in fact associated with the superconductivity.

Here, we have observed a clear suppression of $T_\text{TRSB}$ with hydrostatic stress, at a rate matching the suppression of $T_\text{c}$. This result further strengthens the evidence that enhanced muon spin relaxation is an indicator of TRSB superconductivity: $T_\text{TRSB}$ tracks $T_\text{c}$ when tetragonal lattice symmetry is preserved, while the splitting induced by uniaxial pressure shows unambiguously that it is a distinct transition, and not an artefact through some unidentified mechanism of the superconducting transition itself. Figure~\ref{fig:Tc-TTRSB_correlations} shows $T_\text{TRSB}$ versus $T_\text{c}$. The data reported here, on hydrostatically pressurised Sr$_2$RuO$_4$ and on unpressurised Sr$_{1.98}$La$_{0.02}$RuO$_4$, fall on the $T_\text{TRSB} = T_\text{c}$ line, while the uniaxial pressure data from Ref.~\onlinecite{Grinenko_SRO_NatPhys_2020} clearly deviate from this line.

Our central finding that $T_\text{TRSB}$ tracks $T_\text{c}$ provides further support for the single-representation $d_{xz} \pm id_{yz}$ order parameter. Importantly, $d_{xz} \pm id_{yz}$ is the only spin-singlet order parameter consistent with the selection rules imposed by ultrasound and Kerr effect data.  The sound velocity for longitudinal ultrasound modes is renormalized at a superconducting transition, generally. A jump at $T_\text{c}$ in ultrasound velocity for transverse modes, however, is a signature of a multi-component order parameter.  Ultrasound data on Sr$_2$RuO$_4$ show precisely this type of renormalization \cite{Ghosh_NatPhys_2020, Benhabib_NatPhys_2020}. While these experimental results are not sensitive to the spin configuration, they impose other stringent conditions on the possible pairing symmetries \cite{Walker_PRB_2002, Sigrist_ProgTheorPhys_2002}.  The polar Kerr effect mentioned above is a second experiment which provides symmetry-related constraints, being compatible only with chiral pairing states \cite{Xia_PRL_2006}. These two selection rules are obeyed by both the chiral $p$-wave and chiral $d$-wave state, though as noted in the Introduction, $p$-wave order appears to be ruled out by NMR Knight shift data \cite{Pustogow_Nature_2019,Ishida_JPSJ_2020}.  In contrast, the composite-representation states do not satisfy the requirements for both selection rules. The $d_{x^2-y^2}+ig_{xy(x^2-y^2)}$ and $s+id_{xy}$ states are constructed to be compatible with the ultrasound measurements, but they are not chiral \cite{Kivelson_npjQuantMat_2020,Romer_arxiv_2021}. The $s+id_{x^2-y^2}$ state violates both selection rules \cite{Romer_PRL_2019}. It can be generally stated that any composite-representation pairing states in a tetragonal crystal, composed of components of two one-dimensional representations, would satisfy at most one of the two selection rules (see the Methods section).

Major challenges to $d_{xz} \pm id_{yz}$ order are the absence of a resolvable second heat capacity anomaly at $T_\text{TRSB}$ in measurements on uniaxially pressurised Sr$_2$RuO$_4$~\cite{Li_arxiv_2019}, and, as already noted, the theoretical challenges in obtaining a horizontal line node in a highly two-dimensional metal~\cite{Zutic_PRL_2005}. We note in addition that an analysis of low-temperature thermal conductivity data indicated vertical, rather than horizontal, line nodes in Sr$_2$RuO$_4$~\cite{Hassinger_PRX_2017}. The theoretical objection to horizontal line nodes may be overcome through the complex nature of the multi-orbital band structure, including sizable spin-orbit coupling~\cite{Suh_PRR_2020, Gingras_PRL_2019, Puetter_EPL_2012}.

So we may conclude that our ZF-$\mu$SR data combined with the selection rules for ultrasound and polar Kerr effect and the NMR Knight shift behavior are consistent with the single-representation chiral $ d_{xz} + i d_{zy} $-wave state, while all composite-representation states suffer from several deficiencies. We note, however, that there are also empirical challenges to a hypothesis of $d_{xz} \pm id_{yz}$, and that the difficulty in reconciling apparently contradictory experimental results in Sr$_2$RuO$_4$ may mean that one or more major, apparently solid results will in time be found to be incorrect, either for a technical reason or in interpretation. Further experiments are therefore necessary.

\section{Methods}
\renewcommand{\theequation}{M\arabic{equation}}
\renewcommand{\thefigure}{ED\arabic{figure}}
\renewcommand{\thetable}{ED\arabic{table}}
\setcounter{equation}{0}
\setcounter{figure}{0}
\setcounter{table}{0}

\subsection{Sample preparation and characterisation}

\subsubsection{Sr$_{\rm 2-y}$La$_{\rm y}$RuO$_4$ single crystals}

Single crystals of Sr$_{\rm 2-y}$La$_{\rm y}$RuO$_4$ were grown by means of a floating zone technique \cite{SRO_sample-grow}. Samples for measurement under hydrostatic pressure (with $y=0$), were cut
from two rods, C140 and C171, that each grew along a $\langle 100 \rangle$ crystallographic direction. The rods have diameter $\varnothing\simeq 3$~mm. Two sections of length 8--12~mm were taken from each rod.
These were then cleaved, forming semi-cylindrical samples with flat surfaces perpendicular to the $c$-axis.

The effect of La doping on the TRSB transition was studied on a single original Sr$_{\rm 2-y}$La$_{\rm y}$RuO$_4$ crystal of length 8~mm. The La concentration was analyzed by an electron-probe micro-analysis and was found to be $y \simeq 0.02$. Before the $\mu$SR measurements, this rod was then cleaved into two semi-cylindrical pieces, again with the flat faces
$\perp c$.

\subsubsection{Specific heat of Sr$_{\rm 2-y}$La$_{\rm y}$RuO$_4$ at ambient pressure}

Specific heat measurements were performed at zero pressure for several pieces of Sr$_{\rm 2-y}$La$_{\rm y}$RuO$_4$ single crystals, cut from the rod used for $\mu$SR measurements.

\begin{figure}[tbh]
\includegraphics[width=1.0\linewidth]{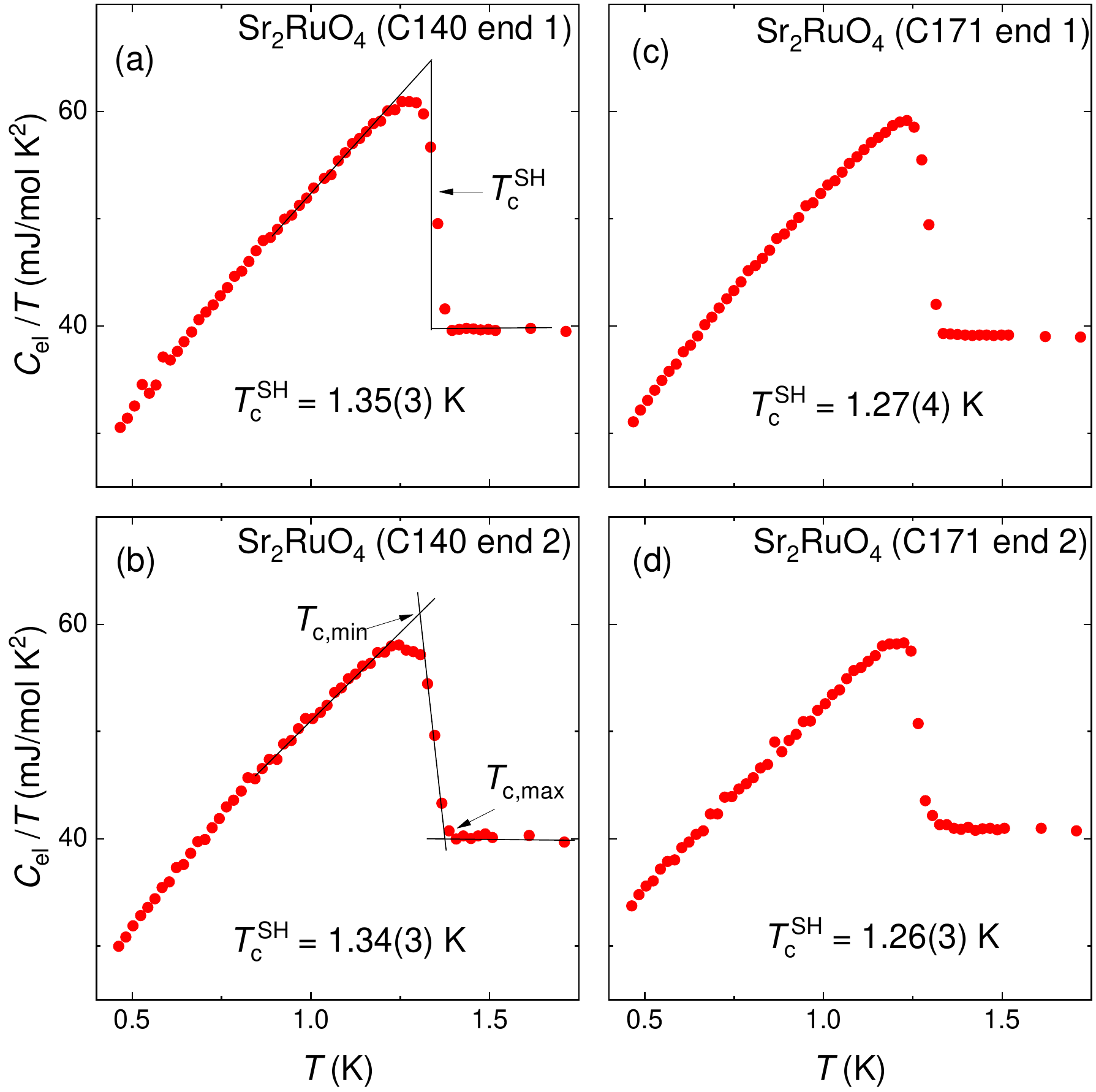}
\caption{ Specific heat curves taken for four ending pieces of C140 and C171 Sr$_2$RuO$_4$ rods. The mean value of the 'specific-heat' superconducting transition temperature $T_{\rm c}^{\rm SH}$ is obtained by an equal-entropy construction of the idealized specific heat jump [panel (a)]. The minim and maximum values of the transition temperature ($T_{\rm c, min}$ and $T_{\rm c, max}$) are determined from the crossing points of linearly extrapolated $C_{\rm el}/T$ vs. $T$ curves in the vicinity of $T_{\rm c}$  [panel (b)].   }
\label{fig:SupplMat_specific-heat_SRO}
\end{figure}

For Sr$_2$RuO$_4$ used in hydrostatic pressure measurements, the electronic specific heat capacity $C_{\rm el}/T$ was measured for four samples: one sample cut from each end of both the C140 and C171 sections. Results are presented in Fig.~\ref{fig:SupplMat_specific-heat_SRO}. The specific-heat critical temperature $T_\text{c}$ of each sample was obtained by an equal-entropy construction, illustrated in panel (a). The spread on the critical temperature of each sample is taken as $T_\text{c, max} - T_\text{c, min}$, where $T_\text{c, max}$ and $T_\text{c, min}$ are determined for each transition as illustrated in panel (b). $T_\text{c}$ was found to be $1.35(3)$ and $1.34(3)$~K for the two samples from rod C140, and $1.27(4)$ and $1.26(3)$ for those from C171. Because both rods were used in the hydrostatic pressure measurements, we take a combined value $T_\text{c}^\text{SH} = 1.30(6)$~K for the specific-heat critical temperature of these samples together.

\begin{figure}[tbh]
\includegraphics[width=0.8\linewidth]{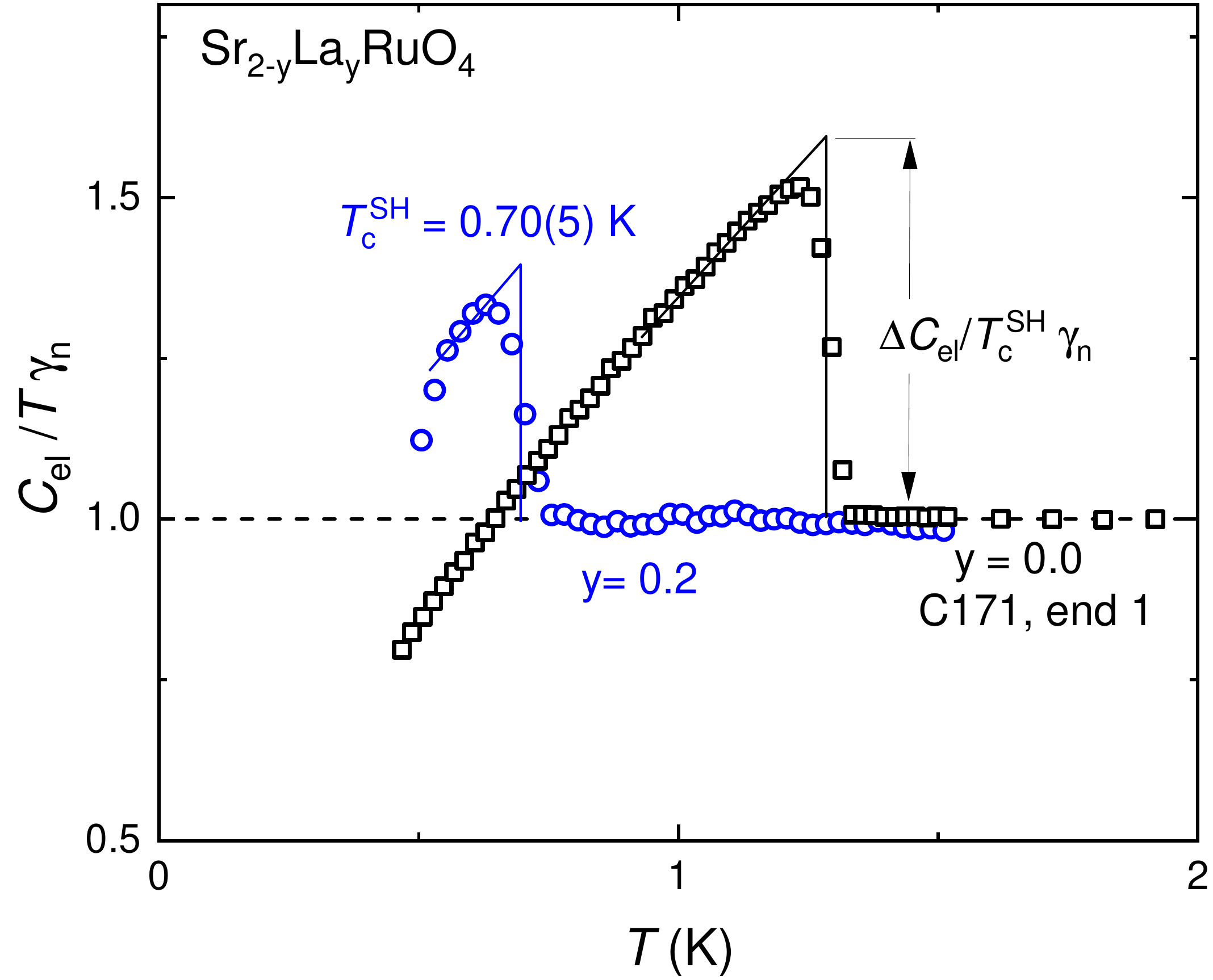}
\caption{ Temperature dependence of the normalized electronic specific  heat $C_{\rm el}/T \gamma_{\rm n}$ measured on a small piece cut from the Sr$_{\rm 1.98}$La$_{\rm 0.02}$RuO$_4$ $\mu$SR sample (blue open circles). Open squares correspond to the $C_{\rm el}/T \gamma_{\rm n}$ vs. $T$ data for one of Sr$_2$RuO$_4$ samples [C171, end 1;  Fig.~\ref{fig:SupplMat_specific-heat_SRO}(c)]. Solid lines represent an equal-entropy construction used to determine the superconducting transition temperature $T_{\rm c}$. The double-sided arrow represent the way of determination of the specific heat jump at $T_{\rm c}$ ($\Delta C_{\rm el}/T_{\rm c} \gamma_{\rm n}$).}
\label{fig:SupplMat_specific-heat_LaSRO}
\end{figure}

The temperature dependence of $C_{\rm el}/T$  for a small piece cut from the Sr$_{\rm 1.98}$La$_{\rm 0.02}$RuO$_4$ $\mu$SR sample is presented in Figs.~\ref{fig:La-doped}~(b) and (e).
Figure~\ref{fig:SupplMat_specific-heat_LaSRO} show the same data, but with $C_{\rm el}/T$ normalised by the Sommerfeld coefficient $\gamma_{\rm n}$. The equal-entropy construction and estimates of
$T_{\rm c, min}$ and $T_{\rm c, max}$ result in $T_{\rm c}^{\rm SH}=0.70(5)$~K.


\subsection{$\mu$SR experiments}
The muon spin rotation/relaxation ($\mu$SR) experiments were performed at the $\mu$E1 and $\pi$E1 beamlines, using the GPD \cite{Khasanov_HPR_2016}, and Dolly spectrometers (Paul Scherrer Institute, PSI
Villigen,  Switzerland). At the GPD instrument, experiments under pressure up to $p\simeq 0.95$~GPa on undoped Sr$_2$RuO$_4$ were performed. At the Dolly spectrometer, measurements of Sr$_{\rm 1.98}$La$_{\rm
0.02}$RuO$_4$ at ambient pressure were conducted. At both instruments $^{4}$He cryostats equipped with the $^{3}$He insets (base temperature $T\simeq0.25$~K) were used.

At the GPD instrument, measurements in zero-field (ZF-$\mu$SR) and with the field applied transverse to the initial muon spin polarization ${\bf P}_{\mu}(0)$ (TF-$\mu$SR) were performed. In two sets
of ZF-$\mu$SR studies, ${\bf P}_{\mu}(0)$ was set to be parallel to the $c$-axis and along the $ab$-plane, respectively. In TF-$\mu$SR measurements the small 3~mT magnetic field was applied parallel
to the $c$-axis and perpendicular to ${\bf P}_{\mu}(0)$.

At the Dolly instrument, in addition to ZF- and TF-$\mu$SR experiments, the longitudinal-field (LF) measurements were performed. In these studies 3~mT magnetic field was applied parallel to the
$c$-axis and to the initial muon spin polarisation ${\bf P}_{\mu}(0)$.

\subsubsection{$\mu$SR data analysis procedure}

The experimental data were analyzed by separating the $\mu$SR signal on the sample (s) and the background (bg) contributions \cite{Khasanov_SciRep_2015}:
 \begin{equation}
A_0  P(t) = A_{\rm s} P_{\rm s}(t) + A_{\rm bg} P_{\rm bg}(t).
 \label{eq:P(t)}
\end{equation}
Here $A_0$ is the initial asymmetry of the muon spin ensemble, and  $A_{\rm s}$ ($A_{\rm bg}$) and $P_{\rm s}(t)$ [$P_{\rm bg}(t)$] are the asymmetry  and the time evolution of the muon spin polarization for muons stopped inside the sample (outside of the sample), respectively.

In a case of $\mu$SR under pressure studies, the background contribution (approximately 50\% of total $\mu$SR response)
is determined by the muons stopped in the pressure cell body. At ambient pressure experiment the small background contribution (of the order of 5\%) is
caused by muons stopped in the sample holder and the cryostat windows.

\subsubsection{TF-$\mu$SR}

In TF-$\mu$SR experiments, the sample contribution was analyzed by using the following functional form:
\begin{equation}
P_{\rm s}^{\rm TF}(t) = \exp\left[ -\frac{\sigma^2t^2}{2} \right]
\cos(\gamma_\mu B_{\rm int}t+\phi).
 \label{eq:TF-P_sample}
\end{equation}
Here $B_{\rm int}$ is the internal field in the sample, $\phi$ is the initial phase of the muon spin ensemble, and $\gamma_{\rm \mu}\simeq 2\pi \times 135.5$~MHz/T is the muon gyromagnetic ratio. The Gaussian relaxation rate $\sigma$ consists of the 'superconducting', $\sigma_{\rm sc}$, and nuclear moment, $\sigma_{\rm nm}$, contributions and it is defined as: $\sigma^2=\sigma_{\rm sc}^2+\sigma_{\rm nm}^2$. Here, $\sigma_{\rm sc}$ and $\sigma_{\rm nm}$ characterize the damping due to the formation of the flux-line lattice in the superconducting state and of the nuclear magnetic dipolar contribution, respectively. In the analysis, $\sigma_{\rm nm}$ was assumed to be constant over the entire temperature range and was fixed to the value obtained above $T_{\rm c}$, where only nuclear magnetic moments contribute to the muon depolarization rate [see Fig.~\ref{fig:p0p25_Tc-TRSB}(a)].

The pressure cell contribution was described by using the following equation:
\begin{equation}
P_{\rm pc}^{\rm TF}(t) = \exp\left[ -\frac{\sigma_{\rm pc}^2t^2}{2} \right]
\cos(\gamma_\mu B_{\rm ext}t+\phi).
 \label{eq:TF-P_pc}
\end{equation}
Here $\sigma_{\rm pc}\simeq 0.28$~$\mu{\rm s}^{-1}$ is the field and the temperature independent relaxation
rate of beryllium-copper  \cite{Khasanov_HPR_2016}, and $B_{\rm ext}$ is the externally applied field.

The solid lines in Fig.~\ref{fig:pressure_Tc-TRSB}(a) correspond to the fit of TF-$\mu$SR data by using Eq.~\ref{eq:P(t)} with the sample and the background parts described by Eqs.~\ref{eq:TF-P_sample} and \ref{eq:TF-P_pc}. For the data presented in Fig.~\ref{fig:La-doped}(a) the background contribution was described by non-relaxing function $P_{\rm bg}^{\rm TF}(t)=\cos(\gamma_\mu B_{\rm ext}t+\phi)$. The good agreement between the fits and the data demonstrates that the above model describes the experimental data  rather well.

With the external magnetic field applied along the crystallographic $c$-axis (${\bf B}_{\rm ext} \parallel c$),
the superconductig contribution into the Gaussian relaxation rate $\sigma_{\rm sc}$
becomes proportional to the inverse squared in-plane magnetic penetration depth
$\lambda_{\rm ab}$ \cite{Khasanov_PRB_2016}. The proportionality coefficient between $\sigma_{\rm sc}$
and $\lambda_{ab}^{-2}$ depends on the value of the applied field, the symmetry of the flux-line lattice and the angular dependence of the
superconducting order parameter.

The temperature dependencies of the Gaussian relaxation rate $\sigma$ and the diamagnetic shift $B_{\rm int}-B_{\rm ext}$
are presented in Figs.~\ref{fig:pressure_Tc-TRSB}(b), (c) and  \ref{fig:La-doped}(b), (c) for
Sr$_2$RuO$_4$ and Sr$_{1.98}$La$_{0.02}$RuO$_4$ samples, respectively.

\subsubsection{ZF-and LF-$\mu$SR}

The sample contribution includes both, the nuclear moment relaxation and an additional exponential relaxation $\lambda$ caused by appearance of spontaneous
magnetic fields \cite{Luke_Nature_1998}:
\begin{equation}
P_{\rm s}^{\rm ZF}(t) = {\rm GKT}_{\rm s}(t)\; e^{-\lambda t}.
 \label{eq:ZF-P_sample}
\end{equation}
Here ${\rm GKT}(t)$ is the Gaussian Kubo-Toyabe (GKT) relaxation function describing
the magnetic field distribution created by the nuclear magnetic moments \cite{Hayano_PRB_1979, Yaouanc_book_2011}:
 \begin{equation}
{\rm GKT}(t) = \frac{1}{3}+\frac{2}{3}(1-\sigma_{\rm GKT}^2t^2)\; e^{-\sigma_{\rm GKT}^2 t^2/2}.
 \label{eq:GKT}
\end{equation}
$\sigma_{\rm GKT}$ is the GKT relaxation rate.

Muons implanted in beryllium-copper pressure cell body sense solely the magnetic field distribution created by copper nuclear magnetic moments and described as:
 \begin{equation}
P_{\rm pc}^{\rm ZF}(t) = {\rm GKT}_{\rm pc}(t)
 \label{eq:ZF-P_GKT}
\end{equation}
with the temperature independent relaxation rate $\sigma_{\rm GKT, BeCu}\simeq 0.35$~$\mu {\rm s}^{-1}$ \cite{Khasanov_HPR_2016}.

Fits of Eq.~\ref{eq:P(t)}, with the sample and pressure cell parts described by Eqs.~\ref{eq:ZF-P_sample} and \ref{eq:ZF-P_GKT}, to the ZF-$\mu$SR data were performed globally. The ZF-$\mu$SR time-spectra taken at each particular muon spin polarization [${\bf P}_\mu(0) \| ab$ and ${\bf P}_\mu(0) \| c$] and pressure ($p=0.0$, 0.25, 0.62,
and 0.95~GPa) were fitted simultaneously with  $A_{\rm s}$, $A_{\rm pc}$, $\sigma_{\rm GKT, Sr_2RuO_4}$, $\sigma_{\rm GKT, BeCu}$, and $\lambda_0$ as common parameters, and $\lambda$ as individual parameter for each particular data set.
The solid green and purple lines in Figs. \ref{fig:pressure_Tc-TRSB}~(d)
correspond to the fit of ZF-$\mu$SR data by using Eq.~\ref{eq:P(t)} with the sample and the background parts
described by Eqs.~\ref{eq:ZF-P_sample} and \ref{eq:ZF-P_GKT}.

Note that the absence of strong nuclear magnetic moments in Sr$_{2-y}$La$_{y}$RuO$_4$ leads to the corresponding Gaussian Kubo-Toyabe relaxation rate being nearly zero. Consequently, the analysis of ZF- and LF-$\mu$SR data for Sr$_{1.98}$La$_{0.02}$RuO$_4$ was performed by using
the simple-exponential decay function:
\begin{equation}
P_{\rm s}^{\rm ZF,LF}(t) = e^{-\lambda t}.
 \label{eq:LF-P_sample}
\end{equation}
The solid lines in Figs. \ref{fig:La-doped}~(d) correspond to the fit of ZF-$\mu$SR data by using Eq.~\ref{eq:P(t)} with the sample part
described by Eqs.~\ref{eq:LF-P_sample} and the non-relaxing background $P_{\rm bg}^{\rm ZF,LF}(t)=1$.

The temperature dependencies of the exponential relaxation rate $\lambda$ are presented in Figs.~\ref{fig:pressure_Tc-TRSB}(e,f) and  \ref{fig:La-doped}(e) for
Sr$_2$RuO$_4$ and Sr$_{1.98}$La$_{0.02}$RuO$_4$ samples, respectively.

\subsubsection{ZF- and TF-$\mu$SR results at $p=0.25$ and $p=0.62$~GPa}

\begin{figure}[tbh]
\includegraphics[width=0.8\linewidth]{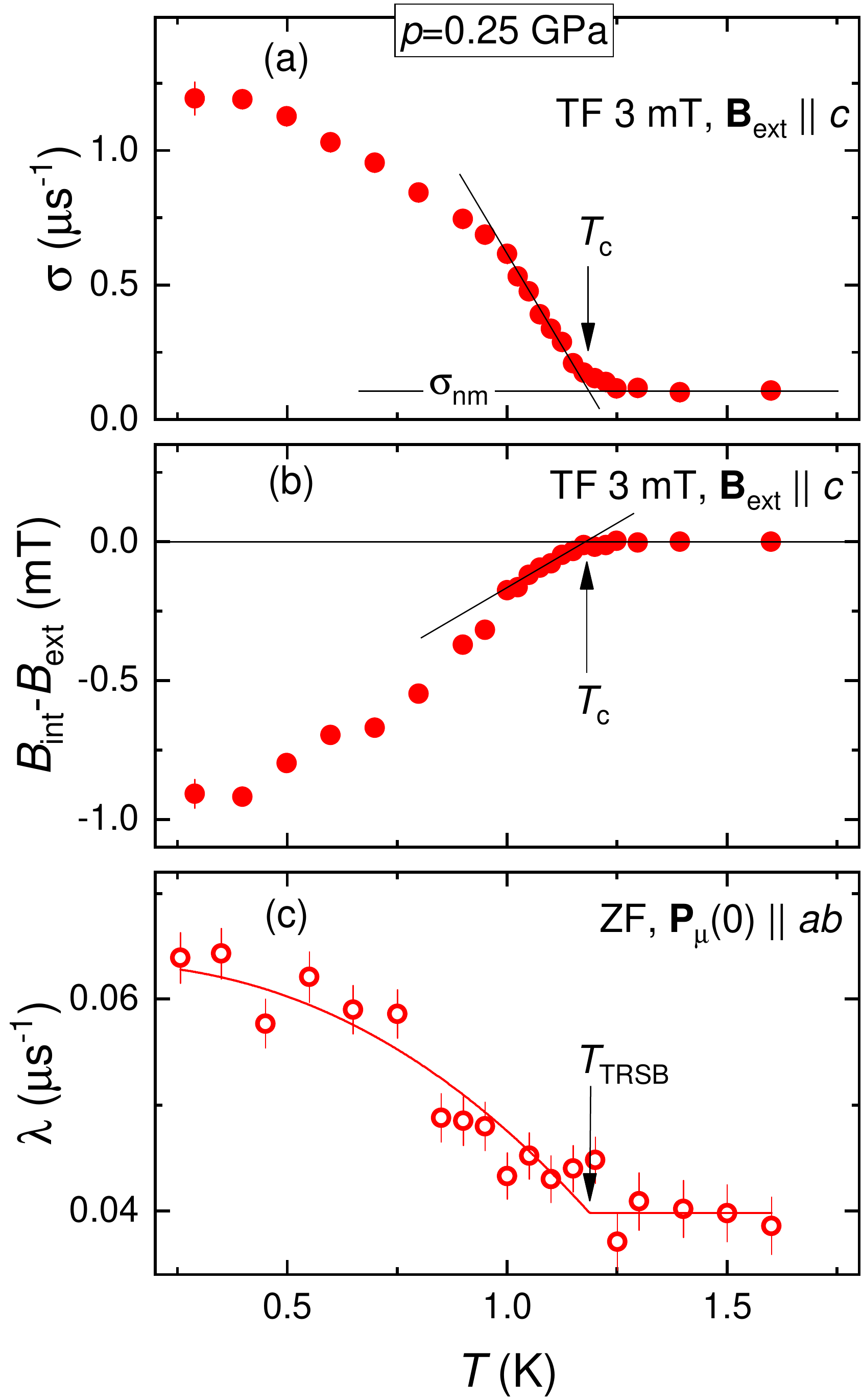}
\caption{(a) Temperature dependence of the Gaussian relaxation rate $\sigma$ measured at $p=0.25$~GPa and the external field $B_{\rm ext}=3$~mT applied parallel to the crystallographic $c$-axis. (b) The diamagnetic shift of the internal field $B_{\rm int}-B_{\rm ext}\propto M_{\rm FC}$ \cite{Weber_PRB_1993}, $M_{\rm FC}$ is the field-cooled magnetization, at $p=0.25$ ~GPa. Arrows in panels (a) and (b) indicate the position of the superconducting transition temperature $T_{\rm c}$. (c) Temperature dependence of the ZF exponential relaxation rate $\lambda$ induced by spontaneous magnetic fields caused by TRSB effects at $p=0.25$~GPa. The initial muon spin polarization ${\bf P}_{\rm \mu}(0)$ is parallel to the $ab$-plane. The solid line is the fit by means of Eq.~\ref{eq:lambda} from the main text. Arrow indicate the position of TRSB transition temperature $T_{\rm TRSB}$.}
\label{fig:p0p25_Tc-TRSB}
\end{figure}

\begin{figure}[tbh]
\includegraphics[width=0.8\linewidth]{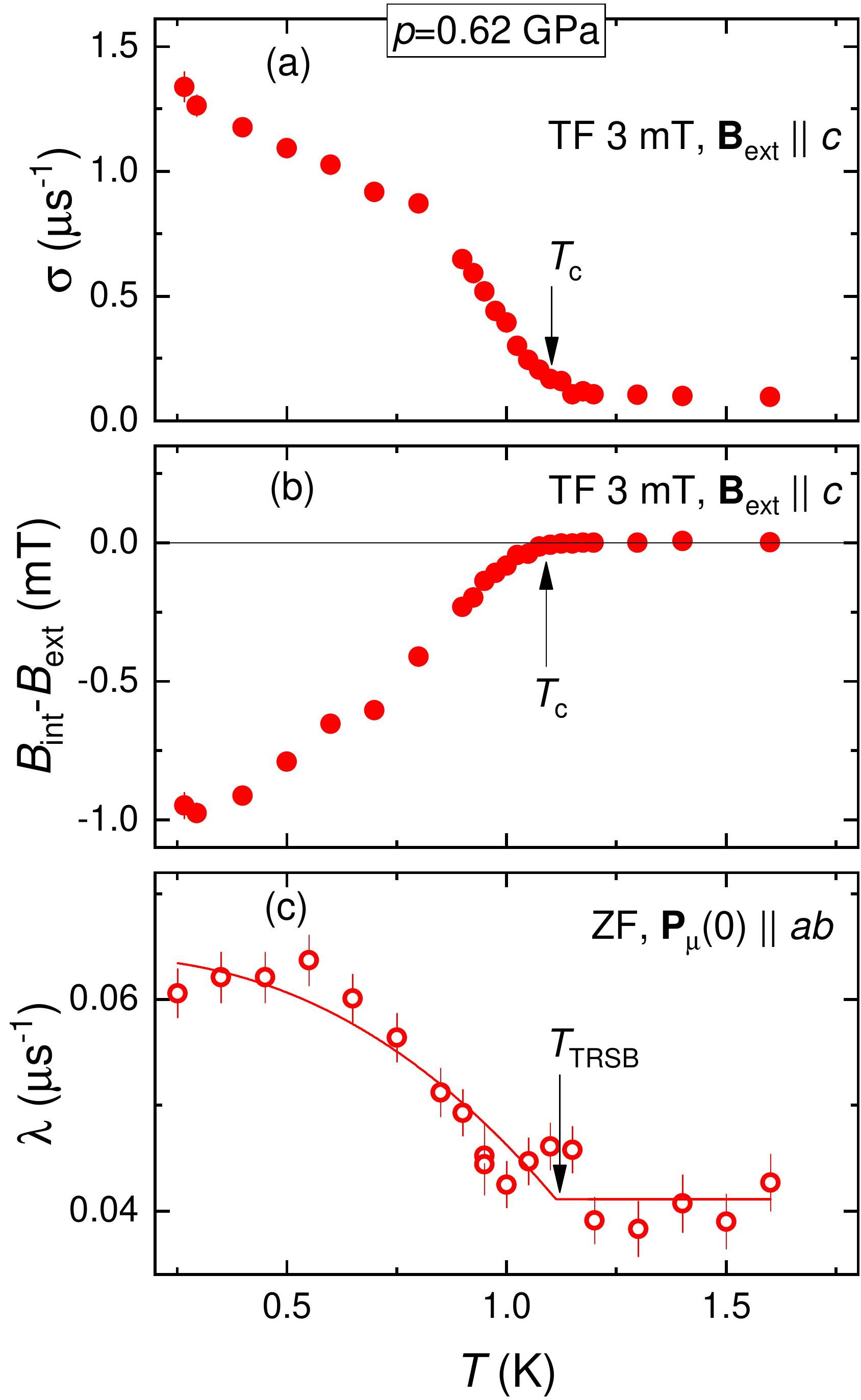}
\caption{The same as in Fig.~\ref{fig:p0p25_Tc-TRSB} but for $p=0.62$~GPa.}
\label{fig:p0p62_Tc-TRSB}
\end{figure}

Figures \ref{fig:p0p25_Tc-TRSB} and \ref{fig:p0p62_Tc-TRSB} show the results of TF- and ZF-$\mu$SR measurements on Sr$_2$RuO$_4$ at $p=0.25$ and 0.62~GPa. Arrows in panels (a) and (b) indicate the position of the superconducting transition temperature $T_{\rm c}$. Arrow in panel (c) indicate the TRSB transition temperature $T_{\rm TRSB}$.

\subsubsection{Extraction of $T_{\rm c}$ from the TF-$\mu$SR data}

The superconducting transition temperature $T_{\rm c}$ was extracted from temperature dependencies of the Gaussian relaxation rate, $\sigma$, and the diamagnetic shift of the internal field, $B_{\rm int}-B_{\rm ext}\propto M_{\rm FC}$, as they obtained in TF-$\mu$SR experiments [see Figs.~\ref{fig:pressure_Tc-TRSB}(b,c), \ref{fig:La-doped}(b,c), \ref{fig:p0p25_Tc-TRSB}(a,b), and \ref{fig:p0p62_Tc-TRSB}(a,b)].

In a case of $\sigma(T)$ data, the transition temperature was defined as a crossing point of linearly extrapolated $\sigma(T)$ curve in the vicinity of $T_{\rm c}$ with  $\sigma=\sigma_{\rm nm}$ line. Note that $\sigma_{\rm nm}$ is constant over the entire temperature range and it corresponds to the value reached above $T_{\rm c}$ [see Fig.~\ref{fig:p0p25_Tc-TRSB}(a)].

From the diamagnetic shift data, the transition temperature was defined as a crossing point of linearly extrapolated $B_{\rm int}-B_{\rm ext}$ vs. $T$ curve in the vicinity of $T_{\rm c}$ with  $B_{\rm int}-B_{\rm ext}=0$ line [see Fig.~\ref{fig:p0p25_Tc-TRSB}(b)].

\subsection{Symmetry properties of the order parameters}

Several order parameters have been proposed for the time reversal symmetry-breaking superconducting state of Sr$_2$RuO$_4$.
We would like here to give a brief overview on the different options and the symmetry requirements to satisfy the selection rules
for two experiments: ultrasound velocity renormalization for the transverse $c_{66} $-mode and the polar Kerr effect.
For tetragonal crystal symmetry with the point group $D_{4h} $ the even-parity spin-singlet pairing states can be listed
according to the irreducible representations of $ D_{4h} $, four one-dimensional ones $ A_{1g}, A_{2g}, B_{1g}, B_{2g} $ and
a two-dimensional one $ E_u $. The pair wave function $ \psi_{\Gamma} (\bm{k}) $ of the corresponding states are given by

\begin{equation} \begin{array}{lc}
\psi_{A_{1g}} (\bm{k}) = \psi_0(\bm{k})  &  s\mbox{-wave} \\[2mm]
\psi_{A_{2g}} (\bm{k}) =   \psi_0(\bm{k})k_x k_y (k_x^2 - k_y^2) \quad  &  g_{xy(x^2-y^2)}\mbox{-wave} \\[2mm]
\psi_{B_{1g}} (\bm{k}) =   \psi_0(\bm{k})k_x^2 - k_y^2 &  d_{x^2-y^2}\mbox{-wave} \\[2mm]
\psi_{B_{2g}} (\bm{k}) =   \psi_0(\bm{k})k_x k_y  &  d_{xy}\mbox{-wave} \\[2mm]
\psi_{E_{g}}   (\bm{k}) = \{  \psi_0(\bm{k})k_xk_z, \psi_0(\bm{k})k_y k_z \} & \{d_{xz},d_{yz}\}\mbox{-wave}
\end{array}
\end{equation}
where $ \psi_0(\bm{k}) $ is a function of $ \bm{k} $ invariant under all symmetry operations of the tetragonal lattice. We list here first the composite-representation TRSB states:

\begin{equation}\begin{array}{lr}
\tilde{\Gamma}_1 = A_{1g} \oplus A_{2g} : & {s + i g}\mbox{-wave} \\[2mm]
\tilde{\Gamma}_2 = A_{1g} \oplus B_{1g} : & {s + i d}\mbox{-wave}  \\[2mm]
\tilde{\Gamma}_3 = A_{1g} \oplus B_{2g} : & {s + i d'} \mbox{-wave} \\[2mm]
\tilde{\Gamma}_4 = B_{1g} \oplus A_{2g} : & {d+ i g}\mbox{-wave}  \\[2mm]
\tilde{\Gamma}_5 = B_{2g} \oplus A_{2g} : & {d'+ i g}\mbox{-wave}  \\[2mm]
\tilde{\Gamma}_6 = B_{1g} \oplus B_{2g} : &{d + i d'}\mbox{-wave}  \\[2mm]
\end{array}
\end{equation}
Note that in general different representations correspond to different critical temperature. Thus to obtain a single superconducting phase transition
for the composite states an accidential degeneracy of two representations is necessary.
The two states proposed so far are $ \tilde{\Gamma}_2 $ \cite{Romer_PRL_2019, Romer_PRB_2020} and $  \tilde{\Gamma}_4 $ \cite{Kivelson_npjQuantMat_2020, Willa_PRB_2020}. The two-dimensional
representation allows for the combination
\begin{equation}
\tilde{\Gamma}_7 = E_g : \mbox{chiral }d\mbox{-wave}
\end{equation}
with a pair wave function $ \psi_{E_g} (\bm{k}) = \psi_0(\bm{k}) k_z (k_x \pm i k_y) $ as proposed in Refs.~\onlinecite{Zutic_PRL_2005, Suh_PRR_2020}. All composite states, $ \tilde{\Gamma}_{1-6} $,
can be constructed by electron pairing within the RuO$_2$ planes, while the state $ \tilde{\Gamma}_7 $ requires interlayer pairing. Due to the spin singlet nature all states are compatible with the
new NMR Knight shift results \cite{Pustogow_Nature_2019, Ishida_JPSJ_2020}. All TRSB state are expected to generate internal spontaneous currents around defects, such as surfaces and domain walls and,
consequently, under present understanding are compatible with the $\mu$SR experiments \cite{Luke_Nature_1998}.

Next we consider the two selection rules. For the coupling to the lattice we restrict consideration to the mode which corresponds to the elastic constant $ c_{66} $, which is connected with the strain tensor element $ \epsilon_{xy} = \epsilon_{yx} $ \cite{Walker_PRB_2002, Sigrist_ProgTheorPhys_2002}. This is active for transverse modes with a wave vector in the plane, e.g. [100] and a polarization perpendicular also within the plane. This strain tensor component belongs by symmetry to the representation $B_{2g}$ \cite{Walker_PRB_2002, Sigrist_ProgTheorPhys_2002, Sigrist_RMP_1991}. For the observed renormalization of the speed of sound the superconducting order parameter has to couple linearly to $ \epsilon_{xy} $, thus, requiring that
$ B_{2g} $ is contained in the decomposition of $ \tilde{\Gamma}_j \otimes \tilde{\Gamma}_j $. This only possible for $ \tilde{\Gamma}_3,  \tilde{\Gamma}_4$ and $  \tilde{\Gamma}_7 $:
\begin{equation}
\tilde{\Gamma}_3 \otimes \tilde{\Gamma}_3 = \tilde{\Gamma}_4  \otimes \tilde{\Gamma}_4 = 2 A_{1g} \oplus 2 B_{2g}
\end{equation}
and
\begin{equation}
\tilde{\Gamma}_7 \otimes \tilde{\Gamma}_7 =  A_{1g} \oplus A_{2g} \oplus B_{1g} \oplus B_{2g} .
\end{equation}

The selection rule resulting in the polar Kerr effect requires the order parameter to couple by symmetry to the $z$-component of the magnetic field, $ B_z $ which belongs to the representation $ A_{2g} $. Again we consider the decomposition of the corresponding representations of the different pairing states. We find that only $ \tilde{\Gamma}_1, \tilde{\Gamma}_6 $ and $  \tilde{\Gamma}_7 $ satisfy the condition. The only pairing state which appears to  obey both selection rules is the chiral $d$-wave state. None of the composite pairing states can satisfy both conditions. Among them there are the states
$ \tilde{\Gamma}_2 $ and $ \tilde{\Gamma}_5 $ which are in conflict with both selection rules.

Turning to the odd-parity states the analogous picture arises with
\begin{equation} \begin{array}{ll}
\bm{d}_{A_{1u}} (\bm{k})  &= \psi_0(\bm{k}) (\hat{\bm{x}} k_x + \hat{\bm{y}} k_y) \\[2mm]
\bm{d}_{A_{2u}} (\bm{k})  &=   \psi_0(\bm{k}) (\hat{\bm{x}} k_y - \hat{\bm{y}} k_x) \\[2mm]
\bm{d}_{B_{1u}} (\bm{k})  &=   \psi_0(\bm{k}) (\hat{\bm{x}} k_x - \hat{\bm{y}} k_y)  \\[2mm]
\bm{d}_{B_{2u}} (\bm{k})  &=   \psi_0(\bm{k})(\hat{\bm{x}} k_y + \hat{\bm{y}} k_x)   \\[2mm]
\bm{d}_{E_{u}}   (\bm{k})  &=  \psi_0(\bm{k})  \{\hat{\bm{z}}k_x,\hat{\bm{z}}k_y \}  .
\end{array}
\end{equation}
here listed in the convenient $ \bm{d} $-vector notation for spin-triplet pairing states (see \cite{Sigrist_RMP_1991}).
It is important to note that all composite phases from combination of two pairing states of one-dimensional representation
are $c$-axis equal spin state and would be in agreement with present time NMR Knight data \cite{Pustogow_Nature_2019, Ishida_JPSJ_2020} and had been
proposed as possible states in Refs.~\onlinecite{Kashiwaya_PRB_2019, Ikegaya_PRB_2020}. These states are also called helical state in literature, as they are topologically non-trivial with helical surface states.
The Knight shift experiments disagree with expectations of the state in representation $ E_u $ which yields the chiral $p$-wave state.

Again we have to make composite states of the one-dimensional representation to obtain TRSB phases. Analogous to the even-parity case we
do not find any composite state which satisfies both selection rules, in contrast to the chiral $p$-wave state which behaves the same way as the chiral $d$-wave state in this respect.

\section*{DATA AVAILABILITY}
The data represented in Figs. 2 -- 4 are available as Source Data. All other data that support the plots within this paper and other findings of this study are available from the corresponding author upon reasonable request.

\section*{ACKNOWLEDGMENTS}
The work was performed at the Swiss Muon Source (S$\mu$S), Paul Scherrer Institute (PSI, Switzerland).  We acknowledge fruitful discussions with Zurab Guguchia, Carsten Timm, and Jing Xia. Matthias Elender is acknowledged for technical support.
The work of R.G. is supported by the Swiss National Science Foundation (SNF Grant No. 200021-175935). The work of M.S and B.Z  was financially supported by the Swiss National Science Foundation (SNSF) through Division II (Grant No. 184739). The work of V. G. was supported by DFG GR 4667/1. N.K. acknowledges the support from JSPS KAKENHI (No. JP18K04715) in Japan. Y.M. acknowledges funding by JSPS-CNR-SPIN Core-to-core program on oxide superspin project,  and by JSPS KAKENHI Nos. JP15H05852, JP15K21717, and JP17H06136. The work of H.-H.K was supported by DFG SFB 1143 and GRK 1621.

\section*{AUTHOR INFORMATION}
\subsection*{Author contributions}
R.K, V.G, and M.S conceived the project.
Data were taken by R.K, V.G. D.D, and R.G. R.K and V.G performed data analysis and interpreted the results together with M.S.. B.Z. and M.S. provided the theoretical analysis. N.K. provided and characterized samples. R.K., V.G., M.S., and C.W.H. wrote the manuscript with inputs from all authors.
\subsection*{Corresponding Authors}
Correspondence to Vadim Grinenko or Manfred Sigrist or Rustem Khasanov.

\section*{ETHICS DECLARATIONS}
\subsection*{Competing interests}
The authors declare no competing interests.

\end{document}